\pdfoutput=1
\documentclass[conference,9pt]{IEEEtran}
\usepackage{cite}
\usepackage{amsmath,amssymb,amsfonts}
\usepackage{graphicx}
\usepackage{booktabs}
\usepackage[hyphens]{url}

\let\origsmall\small
\renewcommand{\small}{\origsmall\abovedisplayskip=3.5pt plus 2pt minus 1pt\belowdisplayskip=3.5pt plus 2pt minus 1pt\relax}

\graphicspath{{figures/}}

\usepackage[hidelinks]{hyperref}

\begin{document}
\bstctlcite{IEEEexample:BSTcontrol}

\title{Chimera: A Configurable LLM Decode Accelerator Architecture with Compute-Memory-Floorplan Co-Design for HBM-Equipped Multi-Die FPGAs}

\author{\IEEEauthorblockN{Zhiheng Chen, Yifan Zhang, Yu-An Chou, Ye Qiao, Sitao Huang}
\IEEEauthorblockA{University of California, Irvine\\
\{zhihenc5, yifanz58, chouy10, yeq6, sitaoh\}@uci.edu}}

\maketitle

\begin{abstract}
GPUs are widely used for LLM inference,
but single-request decoding is memory-bound,
leaving their compute underutilized while they draw substantial power.
HBM-equipped FPGAs are well suited to LLM decoding:
their customizable datapaths use memory bandwidth
more energy-efficiently than GPUs.
However, high memory bandwidth alone does not ensure high decode performance:
exploiting it requires parallel compute and memory interfaces,
whose resource costs and placement on the chip can limit the throughput.
We present Chimera, a configurable LLM decode accelerator architecture
that co-designs compute, memory access, and floorplan.
Specifically, HBM Proxy Ports extend HBM access to compute lanes on every super logic region (SLR) die,
and tiered expert storage holds mixture-of-experts weights beyond the HBM capacity.
In addition, a unified configuration analysis estimates per-token latency
and per-SLR resource use
and guides the design of each configuration within resource budgets.
On an AMD Alveo U280, we build ten designs for eight dense,
mixture-of-experts, hybrid linear/standard-attention, and ternary LLMs,
up to an 80B-parameter mixture-of-experts LLM on four memory tiers.
Chimera's designs achieve
a geometric mean of 1.6$\times$ the token throughput
of prior state-of-the-art FPGA implementations
and 3.0$\times$ and 1.5$\times$ the energy efficiency
of the RTX 3090 and A100 GPUs,
setting a new benchmark for LLM decoding on HBM-equipped FPGAs.
A demo is available at \url{https://anonymous.4open.science/r/chimera_fpga_llm_demo-0658/}.

\end{abstract}

\section{Introduction}
\label{sec:intro}

GPUs are the predominant platform for LLM inference~\cite{kwon2023vllm},
but autoregressive decoding rereads the LLM parameters and runtime state,
such as the KV cache, for every token,
making effective memory bandwidth a key performance bottleneck~\cite{zeng2024flightllm,li2025pushing,qiao2026tellmev2,zhang2026pdswap}.
At batch size one, decoding is memory-bound,
leaving GPU compute underutilized~\cite{zeng2024flightllm,he2025lutllm}
while the GPU draws substantial power.
In contrast, FPGAs equipped with high-bandwidth memory (HBM) support customized datapaths
with low-bit arithmetic, parallel channel access, and intermediate results kept on chip,
which use bandwidth efficiently at lower power than GPU execution~\cite{zeng2024flightllm,zhang2026flexllm}.
They provide less peak bandwidth than GPUs,
49\% and 30\% of that of the RTX 3090 and A100 (Table~\ref{tab:platforms}),
and fewer compute resources~\cite{he2025lutllm},
but this efficiency makes them attractive platforms for LLM decoding.

\begin{figure}[t]
  \centering
  \includegraphics[width=\linewidth]{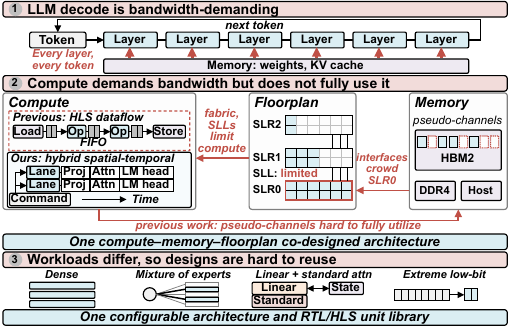}
  \caption{Challenges of LLM decode acceleration on HBM-equipped FPGAs and our approach. SLR: super logic region, one die. SLL: super long line, a wire between dies.}
  \label{fig:challenges}
\end{figure}

However, \textbf{available HBM bandwidth alone does not ensure
high decode performance}.
Using more bandwidth requires compute logic to consume the data
and interfaces to deliver it, and both occupy logic, buffers, and routing.
On a multi-die FPGA, these costs are unevenly distributed:
regions near HBM hold most of the memory-interface logic, while
compute placed farther away requires additional wires between dies.
A design can therefore run out of the resources of one die before
exhausting either its arithmetic capacity or nominal bandwidth.
\textbf{Compute allocation, memory organization, and floorplanning
jointly determine the performance that can be realized.}
Figure~\ref{fig:challenges} summarizes these challenges and our approach.

Architectural and implementation choices change
how compute, interfaces, and routing share these costs.
Spatial dataflow gives each operator its own module,
but decode dependencies and limited bandwidth
leave some modules underutilized~\cite{chen2024spatialllm,zhang2026flexllm}.
Temporal multiplexing shares engines across operators
at the cost of additional control and data routing~\cite{chen2024spatialllm,zhang2026flexllm}.
The overhead of HLS datapaths and interfaces takes resources
that could otherwise add parallelism~\cite{lahti2019hls,guo2023tapa}.
TAPA floorplans and pipelines a given task graph across dies~\cite{guo2023tapa},
but its tasks keep the resource overhead of HLS,
and the designer still chooses the task graph.
Each implementation takes hours of place and route,
so exploring these choices by building each design is costly.

Specialized FPGA LLM accelerators demonstrate efficient
architectures and mapping flows, yet their design choices
remain closely tied to the target
architecture and workload~\cite{zeng2024flightllm,huang2025edgellm}.
StreamTensor compiles LLMs into spatial dataflow accelerators~\cite{ye2025streamtensor},
which share little hardware across operators,
whereas FlightLLM reuses one instruction-driven matrix engine per core across operators~\cite{zeng2024flightllm}.
FlexLLM builds hybrid spatial--temporal designs from
a composable HLS library and sizes them
with latency and bandwidth models~\cite{zhang2026flexllm},
but these models do not estimate the resources of each die
or the wires between dies.
What is more, prior designs keep weights and the KV cache mainly in HBM
and use DDR for smaller data such as lookup
tables~\cite{zeng2024flightllm},
so the HBM capacity limits the LLM size and context length
they support~\cite{zhang2026flexllm}.
Together, these designs leave a \textbf{gap for an architecture
whose compute parallelism, memory access, and placement across dies
are designed together}.
Such an architecture brings more of the HBM bandwidth to the compute logic,
and the HBM capacity no longer limits the LLM size.
Workloads add their own demands:
mixture-of-experts (MoE) LLMs read only the selected experts, from several memory tiers;
hybrid LLMs mixing linear and standard attention add recurrent-state updates;
and arithmetic formats change memory traffic and hardware cost.

We present \textbf{Chimera}, a configurable LLM decode accelerator architecture
for HBM-equipped multi-die FPGAs
that co-designs compute, memory access, and floorplan
across the super logic region (SLR) dies of the FPGA.
Our contributions are:
\begin{enumerate}
    \item \textbf{Configurable LLM decode accelerator architecture.}
    Chimera runs the operators of every layer
    on command-driven compute lanes,
    and each lane reads its own HBM pseudo-channel.
    Lane modes and added units configure the same architecture
    for dense, MoE, hybrid, and ternary LLMs
    (Sections~\ref{sec:hw-cmd} and~\ref{sec:hw-lanes}).
    \item \textbf{Memory access co-designed with the floorplan.}
    HBM Proxy Ports serve the lanes in SLR1 and SLR2,
    and per-SLR resource budgets set the lanes and compute units of each SLR.
    Tiered expert storage holds MoE weights beyond the HBM capacity
    in DDR4, host memory, and NVMe
    (Sections~\ref{sec:hw-place} and~\ref{sec:hw-tiers}).
    \item \textbf{Unified configuration analysis.}
    We estimate the per-token latency and per-SLR resource use of each configuration
    with one performance and resource model.
    A model-guided workflow uses these estimates to guide the design
    (Section~\ref{sec:model}).
    \item \textbf{Accelerator designs on an AMD Alveo U280.}
    We build ten designs for eight LLMs.
    At a context length of 512, decoding 256 tokens,
    they reach 24.5 to 301.1 tokens/s,
    including 24.5 tokens/s for Qwen3-Next-80B-A3B with FP8 weights,
    an LLM larger than the HBM capacity that runs across heterogeneous memory tiers.
    Against four previous state-of-the-art FPGA dense LLM implementations,
    the geometric mean of the throughput ratios is 1.6$\times$;
    for the extreme low-bit BitNet b1.58-2B,
    the ratio is 4.3$\times$.
    With 49\% of the memory bandwidth of the RTX 3090 and 30\% of that of the A100,
    our designs reach on average 70\% and 52\% of the GPUs' throughput, respectively,
    and, at an average card power of 65~W,
    3.0$\times$ and 1.5$\times$ their energy efficiency
    (Section~\ref{sec:evaluation}).
\end{enumerate}

\section{Background and Related Work}
\label{sec:background}

\textbf{LLM decoding on HBM-equipped FPGAs.}
Inference processes the input in a prefill phase, then
decodes one token at a time through dependent operations:
linear layers reread the weights, and attention
reads and updates the KV cache of earlier
tokens~\cite{zeng2024flightllm}.
Compared with prefill, decode offers less reuse of weights
across tokens and is often more sensitive to memory
bandwidth~\cite{li2025pushing,chen2024spatialllm}.
With less raw compute than GPUs, FPGAs suit decode better than
compute-heavy prefill, and decode is the focus of this work~\cite{chen2024spatialllm,li2025pushing}.
HBM-equipped FPGAs provide many independent memory channels,
called pseudo-channels (PCs), that compute logic reads in parallel~\cite{choi2020hbm}.
The speedup from these channels depends on how well compute parallelism
matches the effective bandwidth of the memory
system~\cite{choi2020hbm,li2025pushing}.

\textbf{Effective bandwidth and physical resource constraints
are coupled.}
Compute logic does not receive the peak HBM bandwidth;
the bandwidth it obtains from each channel depends on channel mapping,
outstanding requests, traffic balance, and the path through
the HBM interconnect.
Large AMD FPGAs are built from several SLRs, and HBM connects to the bottom one,
SLR0~\cite{choi2020hbm}.
Memory interfaces and buffers therefore concentrate in SLR0,
and compute in other SLRs reaches memory through relays across dies.

\textbf{Workload variation changes the design trade-offs.}
Each LLM family places different demands on the same platform
(Figure~\ref{fig:challenges}).
Dense LLMs access weights and a KV cache that grows with the context.
MoE LLMs route each token to a few experts from a large
parameter set, so expert placement determines the transfers between
heterogeneous memory tiers: HBM, on-board DDR4, host memory,
and NVMe, in order of decreasing bandwidth.
Hybrid LLMs replace the growing KV cache of their
linear-attention layers with a recurrent state
that persists across tokens and is updated at each token.

These differences call for one architecture whose lanes, memory paths,
and placement are configured together for each LLM.

\textbf{Related work.}
FPGA LLM accelerators on HBM boards~\cite{zeng2024flightllm,huang2025edgellm,zhang2026flexllm,kalyanasundaram2026pensa,he2025lutllm}
and edge boards~\cite{qiao2026tellmev2,zhang2026pdswap,li2025pushing}
tune their decode datapaths for specific LLMs.
FlexLLM sizes a hybrid spatial--temporal HLS design
with latency and bandwidth models~\cite{zhang2026flexllm}.
Unlike FlightLLM's instruction-driven cores~\cite{zeng2024flightllm}
and FlexLLM's HLS library,
Chimera gives each lane its own pseudo-channel,
places lanes in every SLR within per-SLR budgets,
with HBM Proxy Ports serving those outside SLR0,
and stores expert weights in memory tiers beyond HBM.
Stream-HLS, StreamTensor, and CODO~\cite{basalama2025streamhls,ye2025streamtensor,zhang2026codo}
compile neural networks into spatial dataflow designs,
whereas Chimera reuses its lanes across operators over time.
AutoBridge~\cite{guo2021autobridge} and TAPA~\cite{guo2023tapa}
floorplan and pipeline a given task graph
and leave its choice to the designer;
in Chimera, the configuration defines the task graph:
the lanes of each SLR, their channels,
and the placement of compute units and tensors.
Analytical models~\cite{parashar2019timeloop,kwon2019maestro,wu2022sparseloop,zhang2024llmcompass,chen2024spatialllm}
and template explorers~\cite{huang2021cosa,zhang2020dnnexplorer,ye2020hybriddnn}
estimate and schedule DNN and LLM accelerators;
the unified configuration analysis adds per-SLR resources,
inter-SLR wires, and memory tiers to them.

\section{Configurable Accelerator Architecture}
\label{sec:hardware}

\begin{figure*}[t]
  \centering
  \includegraphics[width=\linewidth]{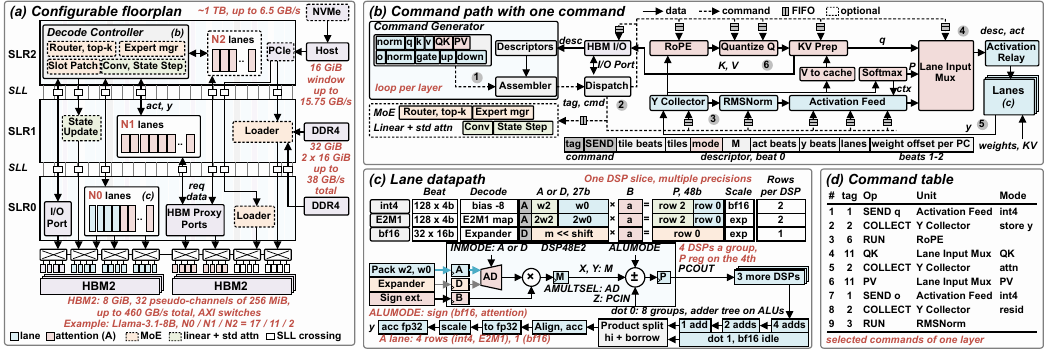}
  \caption{Chimera's configurable accelerator architecture.
  (a) Configurable floorplan, with one example instance; N0, N1, and N2 are the lane counts in SLR0, SLR1, and SLR2.
  (b) Command path, with the format of one command and its descriptor.
  (c) Lane datapath across precisions.
  (d) A command table example: SEND and COLLECT run the lanes in a linear mode, QK and PV switch them to attention, and RUN starts an auxiliary unit.}
  \label{fig:arch}
  \vspace{-11pt}
\end{figure*}

Chimera is a hybrid spatial--temporal architecture for LLM decoding
on an HBM-equipped multi-die FPGA.
Figure~\ref{fig:arch}(a) shows one configuration on the three SLRs,
with HBM2 below SLR0.
A Decode Controller drives every unit with commands (Section~\ref{sec:hw-cmd}),
so the same compute lanes run every operator of every layer over time
(Section~\ref{sec:hw-lanes}).
Three mechanisms tie compute, memory, and floorplan together:
each lane reads its own HBM pseudo-channel,
HBM Proxy Ports in SLR0 serve the lanes of SLR1 and SLR2
(Section~\ref{sec:hw-place}),
and tiered expert storage extends memory beyond HBM
to DDR4, host memory, and NVMe (Section~\ref{sec:hw-tiers}).
A configuration sets the lanes of each SLR,
the optional units a design enables, drawn dashed in the figure,
and the SLR of each unit (Section~\ref{sec:hw-impl}).

\subsection{Command-Driven Execution}
\label{sec:hw-cmd}

Figure~\ref{fig:arch}(b) marks the six steps
in which the Decode Controller runs one layer of one token.
(1)~The Command Generator, through its Assembler stage,
builds the commands of the layer from the per-layer program
and from the operator descriptors that HBM I/O loads
at the start of each token.
(2)~Dispatch writes each command into the queue of the unit its tag names.
Each unit runs its queue in order,
so a command starts when its inputs are ready
and its unit has finished its previous command.
(3)~The Y Collector streams the residual row through RMSNorm
to the Activation Feed, which forms the 8-bit (A8) activation tile.
(4)~On a SEND, the activation tile passes through the Lane Input Mux
and the Activation Relay to the lanes,
and each lane streams its weight tiles from its pseudo-channel.
(5)~On a COLLECT, the Y Collector gathers the lane outputs
and stores them, adds them to the residual row, or applies a gate.
(6)~Attention reuses a subset of the lanes, the attention lanes.
RoPE, Quantize Q, and KV Prep prepare the query,
and HBM I/O appends the new key and value
to the pseudo-channels of the attention lanes.
The QK and PV commands then feed the query and the Softmax probabilities
through the Lane Input Mux,
while the attention lanes read the keys and values from their own channels.
The output projection and the feed-forward network
repeat steps (4) and (5).
Figure~\ref{fig:arch}(d) lists selected commands of one layer in issue order.
An MoE layer adds expert selection,
and its SENDs name expert slots;
a hybrid layer adds the state units of Section~\ref{sec:hw-lanes}.

Decode runs dependent operators one token at a time,
so a module per operator would idle for most of each token.
Temporal reuse instead sizes the lane area
to the parallelism that the bandwidth needs,
and a new LLM family adds units and commands.

\subsection{Compute Lanes}
\label{sec:hw-lanes}

A compute lane is a parallel matrix--vector (GEMV) engine (Figure~\ref{fig:arch}(c)).
An operator's weight tiles, each of whole rows, are dealt round-robin over all lanes.
A lane consumes one 512-bit weight beat per cycle in six pipeline stages.
Weight decode unpacks the beat into the values of the lane's format.
The operand stage places the A8 activation on port B
and the weight on port A,
or on port D for bf16 and FP8.
The dot-product stage has two 32-wide arrays of DSP48E2 slices;
each array chains groups of four slices through the cascade path
and adds the group sums in an adder tree on the DSP ALUs.
In the 4-bit modes, port A packs two weights from different matrix rows,
so one multiplication serves two rows and a beat yields four rows over the two arrays;
the product-split stage separates the sums of the two rows.
8-bit formats yield two rows a beat and bf16 one.
The scale stage aligns each beat's sum to the shared exponent of its scale group
and applies the group or row scale.
The accumulate stage sums the groups of a tile
and emits one output per row.

A lane's mode is the weight format and the operation it runs.
The set of modes a lane supports is a build-time mask over the modes
of the RTL library;
synthesis removes the logic of every mode outside the mask.
At run time, the descriptor of each operator selects one mode of the set,
and the mode changes only between operators.
A design therefore carries only the modes its LLM uses,
and its lanes still switch between operators of different formats.
Figure~\ref{fig:arch}(c) shows one example mode set with three modes:
int4, 4-bit floating point (E2M1) with a shared scale, and bf16.
Other mode sets in the library cover 8-bit integer, FP8, and ternary weights.
With ternary weights, each DSP slice multiplies an A8 value
by three packed weights.

The attention lanes run attention in two more modes on the same DSP slices:
$QK^{\mathsf{T}}$ for the attention scores
and $PV$ for the weighted values.
Keys are stored in blocks of positions,
each block in the pseudo-channel of one attention lane,
and values are stored transposed and split by output dimension over these lanes.
In these modes, the keys or values take the place of the weights,
and the Lane Input Mux feeds the query tiles or the probabilities
in place of an activation tile.
Each design chooses whether attention runs on shared lanes,
which also run the linear operators, or on lanes of its own.
A shared lane uses fewer DSPs than separate GEMV and attention lanes,
which can move attention to another SLR at the cost of more inter-SLR wires.

The Y Collector holds the residual row and two row buffers
and applies bias, residual addition, and gating.
Hybrid LLMs add a Conv unit and two state units.
Each attention lane keeps a copy of the recurrent state of a layer
in its pseudo-channel and multiplies it by the token's vectors
in a QK-mode pass.
A State Step unit in the Decode Controller computes the gates
and the layer output from these products,
and a State Update unit, which holds its own copy of the state,
computes the new state and refreshes the lanes' copies.

\subsection{Channel Access and Placement Across SLRs}
\label{sec:hw-place}

Each lane streams weights from its own HBM2 pseudo-channel
(Figure~\ref{fig:arch}(a)).
The lanes of a design therefore set its parallelism
and the bandwidth it uses together:
each added lane adds the bandwidth of one channel.
A lane keeps many read requests outstanding,
so that its channel streams close to its peak rate.
When several masters read one pseudo-channel,
they share its bandwidth;
one channel per lane gives every lane the whole bandwidth of its channel.

Lane masters outside SLR0 lower the clock of every HBM2 port.
The HBM master of every lane therefore sits in SLR0.
Lanes in SLR1 and SLR2 read through HBM Proxy Ports,
which hold one read-only AXI master in SLR0 for each remote lane.
The remote lane sends its read addresses to its proxy master
and receives the data stream back.
Lanes in every SLR thus read at the full HBM clock rate.
The cost is SLR-crossing wires and relays.
In return, the lanes use the logic of all three SLRs
instead of crowding SLR0, which already holds the memory interfaces.
Section~\ref{sec:eval-guided} measures the port rate with and without Proxy Ports.

The lanes and the other library units are grouped into compute units (CUs),
the Vitis kernel instances.
Each CU is placed in a pblock, a placement region within one SLR,
and each CU port is bound to one pseudo-channel.
Streams between CUs in different SLRs cross on super long lines (SLLs),
the wires between SLRs.
The lanes and CUs of each SLR are therefore configuration parameters:
they decide where the logic, the channel ports, and the SLR crossings fall,
and with them whether a design routes.
Per-SLR resource budgets bound these parameters,
so the floorplan is chosen with the configuration
rather than left to the placer.

\subsection{Tiered Expert Storage}
\label{sec:hw-tiers}

Large MoE LLMs hold more expert weights than the HBM,
while each token selects only a few experts in each layer.
Tiered expert storage keeps the experts in four memory tiers,
in order of decreasing rate:
HBM2, the on-board DDR4,
host memory (reached through a window over PCIe),
and an NVMe drive on the host (Figure~\ref{fig:arch}(a)).
A design whose experts exceed the HBM capacity enables the slower tiers it needs.

The Router, run on the lanes or as its own unit, scores the experts,
and the top-$k$ unit selects the active experts of each token
and their routing weights.
The Expert Manager tracks which experts reside in HBM2 slots.
On a miss, it asks the Loaders to copy the expert into a slot
from DDR4 or from host memory, which the Loaders read over PCIe as bus masters;
host software stages experts from NVMe into host memory.
The slower tiers keep every expert,
so the copy overwrites the least-used slot and writes nothing back.
The Slot Patch, on the command stream of the Activation Feed,
adds the slot of each selected expert to the weight offsets
in the descriptor of its SEND,
so the lanes run the expert as an ordinary GEMV from HBM2.
A missing expert holds only its own SEND until its copy completes.
The Router of the next layer also runs on the current layer's input,
and the Loaders copy the experts it selects into slots
while the lanes compute the current layer.
The Y Collector applies the routing weights,
and a design sums the expert outputs in the Y Collector
or in one merged down projection on the lanes.

Whichever tier holds an expert, the lanes read it from HBM2,
and only the Loaders read the slower tiers.
The number of resident experts is a configuration parameter,
set by the HBM2 capacity that the dense weights and the KV cache leave free.
The capacity of all tiers, not of HBM alone, then bounds the LLM size.

\subsection{Implementation and Configuration}
\label{sec:hw-impl}
Each operator has an HLS C model, which defines its function.
The Decode Controller is a graph of TAPA tasks~\cite{guo2023tapa}
that hold the scheduling and control,
and RTL library units implement the compute datapaths;
any task can be replaced by an RTL unit with the same valid/ready ports.
A configuration fixes the hardware of one design.
It specifies the lane counts and mode sets of each SLR,
whether attention shares lanes,
the grouping of units into CUs,
the HBM channel bindings, the enabled memory tiers, the tensor placement,
and the buffer depths.
Each LLM family adds its own parameters:
the resident expert set and the Loader paths for MoE LLMs,
the state units and the storage of their recurrent state
for hybrid LLMs,
and the lane's mode set for ternary LLMs.
Hardware parameters are fixed for each implementation,
while the command sequence selects the operating modes at run time.

\section{Configuration Analysis and Guided Design}
\label{sec:model}

\begin{figure}[t]
  \centering
  \includegraphics[width=\linewidth]{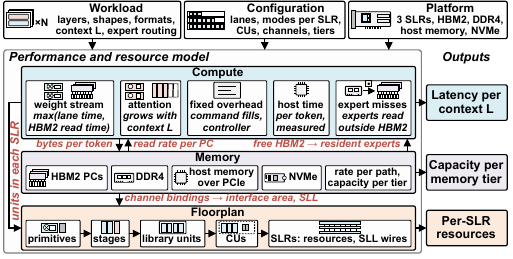}
  \caption{Chimera's performance and resource model.}
  \label{fig:model}
  \vspace{-11pt}
\end{figure}

A configuration of the architecture
fixes compute, memory, and floorplan at once:
each lane takes a pseudo-channel, logic in its SLR,
and wires across SLRs.
This section therefore presents a performance and resource model (Figure~\ref{fig:model})
and a model-guided design workflow,
which together form Chimera's unified configuration analysis.
The model estimates the per-token latency and per-SLR resource use
of a configuration $\mathbf{x}$ for a workload before implementation.
The model is calibrated from measurements.
Unit costs come from synthesis, simulation, and board tests of each library unit;
the memory rates, the controller time, and the host time per token
come from board and microbenchmark runs;
and the resource budgets of Eq.~\eqref{eq:budgets} come from place-and-route runs of earlier prototype versions.

\textbf{Latency.}
The latency of one token at context length $L$ is
\begin{small}
\begin{equation}
T(L) = T_{\mathrm{w}} + T_{\mathrm{attn}}(L) + T_{\mathrm{fix}}
+ \mathbb{E}\left[T_{\mathrm{miss}}(L)\right] + T_{\mathrm{host}}.
\label{eq:time}
\end{equation}
\end{small}
The weight time $T_{\mathrm{w}}$ is the longer of
the lanes' compute time
and the read time of their weights at the measured rate of a pseudo-channel.
$T_{\mathrm{attn}}(L)$ grows linearly with $L$.
$T_{\mathrm{fix}}$ is the command and controller time of each layer
that the lanes do not hide.
$\mathbb{E}[T_{\mathrm{miss}}(L)]$ is the expected extra time of expert misses,
and $T_{\mathrm{host}}$ the host time per token.
Over its contexts $L\in\mathcal{L}$ with shares $w_L$,
a design's throughput is $H(\mathbf{x})=1/\sum_{L} w_L\,T(L;\mathbf{x})$.

\textbf{Memory capacity and expert access.}
The weights, KV cache, and recurrent state placed on
tier or bank $m$, of total size $M_m(L;\mathbf{x})$, stay within its capacity:
$M_m(L;\mathbf{x}) \leq C_m$ at every $L\in\mathcal{L}$.
The model uses the measured rate of each path, not its peak (Table~\ref{tab:platforms}):
with its 128 outstanding requests,
a lane reads 12.6~GB/s from its pseudo-channel, 88\% of the channel's peak.
In an MoE layer, a selected expert that is not resident in HBM2 is a miss,
and a routing profile from sample text gives the miss rate of a resident expert set.
The layer time is the longer of the lane time for the experts
and the Loader time to copy the missed ones.

\textbf{Per-SLR resources and budgets.}\phantomsection\label{sec:model-resources}
Resource use is composed per SLR:
the use $A_{s,r}$ of resource type $r$ in SLR $s$ sums
the platform logic, the memory interfaces,
the units placed in that SLR, and the inter-SLR relays.
SLL demand comes from the stream widths at each SLR crossing.
The model estimates the configurable logic block (CLB) use of an SLR
from its LUT and FF use.
The configuration keeps each SLR within its budgets,
\begin{small}
\begin{equation}
A_{s,r}(\mathbf{x}) \leq \overline{A}_{s,r},
\qquad
D_e(\mathbf{x}) \leq \overline{D}_e,
\qquad
P(\mathbf{x}) \leq \overline{P},
\label{eq:budgets}
\end{equation}
\end{small}
where $D_e$ is the SLL demand at SLR boundary $e$
and $P$ the number of HBM2 ports in use.
The budgets on CLB use, SLR0 LUTs, BRAM, SLL demand, and HBM2 ports
are set below capacity, the last at 31 of the 32 ports;
DSP and URAM are bounded by capacity.

\begin{figure}[t]
  \centering
  \includegraphics[width=\linewidth]{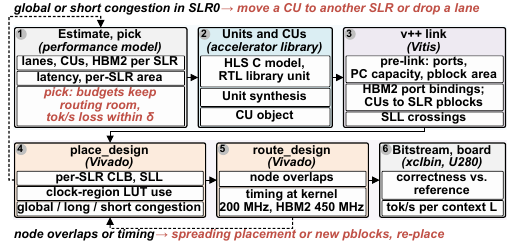}
  \caption{The model-guided design workflow, with the FPGA checks at each step and the feedback from each check to the step it changes.}
  \label{fig:workflow}
  \vspace{-11pt}
\end{figure}

\textbf{Model-guided design workflow.}\phantomsection\label{sec:optimisation}
The workflow (Figure~\ref{fig:workflow}) combines
the model's ranking with checks on the implementation in six steps.
Scripts run every step and check, the same for every LLM family.
After a failed check, the model ranks the possible changes
by the throughput each is estimated to lose,
and the one that loses least is applied.

Step 1 enumerates candidates over
the lane counts N0, N1, and N2 of the three SLRs (Figure~\ref{fig:arch}(a)),
the resident expert set,
the split of weights over the memory tiers, and the controller options.
The model estimates every candidate,
and the chosen configuration, the pick,
is the candidate of highest estimated throughput $H$ within the budgets of Eq.~\eqref{eq:budgets}.
The pick loses at most $\delta$, 10\% for every design here,
against the best candidate without the budgets.
If all candidates within the budgets lose more,
the workflow picks, among the candidates within $\delta$ of that best,
the one closest to the budgets,
and placement and routing decide whether it is kept.

After step 4, the placer reports global, long, and short congestion levels per SLR,
where levels~5, 6, and 7 are moderate, high, and above high~\cite{amd2023ug949}.
Global or short congestion in SLR0 sends the design back to step 1,
which moves one CU to another SLR or removes one lane from SLR0.
In step 5, node overlaps
(routing resources claimed by more than one net)
or a timing failure send the design back to step 4
for a logic-spreading placement or new pblocks.
Step 6 checks the board output
against the reference LLM and measures tokens/s.

For example, for Llama-3.1-8B with 4-bit weights,
the 31-lane configuration has the highest estimated throughput
but uses every HBM2 port, above the port budget.
With 30 lanes, a 2-lane CU in SLR1 brings the SLR1 CLB use to 92\%,
above its budget,
and the same CU in SLR2 lowers it to 83\%.
The pick is the 30-lane configuration with the CU in SLR2
(Table~\ref{tab:ablation}(b)),
2.5\% below the 31-lane configuration in estimated throughput.

\section{Experiment and Evaluation}\label{sec:evaluation}

\textbf{Experimental setup.}\label{sec:eval-setup}
We build designs for an AMD Alveo U280 (Table~\ref{tab:platforms})
with TAPA~\cite{guo2023tapa}, Vitis and Vivado 2023.1.
Table~\ref{tab:results}(a) lists the ten designs for eight LLMs:
the dense Llama-3.2-1B and Llama-3.1-8B~\cite{hf_llama32_1b,hf_llama31_8b};
the MoE LLMs Qwen1.5-MoE-A2.7B~\cite{hf_qwen15moe} and Qwen3-Coder-30B-A3B~\cite{hf_qwen3coder_fp8};
the hybrid Granite-4.0-H-1B~\cite{hf_granite4h1b}
and the hybrid MoE Qwen3-Next-80B-A3B~\cite{hf_qwen3next_fp8};
and BitNet b1.58-2B~\cite{hf_bitnet2b4t} and the hybrid Ternary-Bonsai-27B~\cite{hf_bonsai27b},
both with ternary weights.
Every design uses A8 activations
and keeps its KV cache and recurrent state in HBM.
Table~\ref{tab:scheme} lists the quantization scheme of each design
with its WikiText-2 perplexity and MMLU score
beside those of the quantized Hugging Face Transformers LLM it implements;
they differ by 0.17\% and 0.3 points on average.
The designs support contexts of up to 4K tokens.
Throughput is the wall-clock rate of decoded tokens on WikiText-2 text,
the median of three rounds,
which differ by under 1\% on the U280.
Card power is the sum of the U280's supply rails,
averaged over a separate decode run.
On the RTX 3090 and A100 GPUs, the baseline is vLLM~\cite{kwon2023vllm}
at batch size one with CUDA graphs and its own operator implementations,
and the same weight bit widths as on the board.
For the three designs of Table~\ref{tab:tiers}(a),
vLLM's optimized expert offload~\cite{vllm2026offload} keeps the experts beyond the GPU memory in host DRAM,
and the GPU reads them over PCIe.
GPU power is the card's energy from NVIDIA's management library
over the same decode window.
\begin{table}[t]
  \centering
  \caption{Platform specifications and FPGA resources.}
  \label{tab:platforms}
  \footnotesize
  \setlength{\tabcolsep}{2.8pt}
  \renewcommand{\arraystretch}{0.8999}
  \setlength{\aboverulesep}{1pt}\setlength{\belowrulesep}{1pt}
  \newcommand{\two}[2]{\begin{tabular}[c]{@{}c@{}}#1\\#2\end{tabular}}
  \resizebox{\columnwidth}{!}{%
  \begin{tabular}{@{}ccccccc@{}}
    \toprule
    & RTX 3090 & A100 & U280 & VCU128 & U50 & V80 \\
    \midrule
    Memory & \two{24 GiB}{GDDR6X} & \two{40 / 80 GiB}{HBM2e} & \two{8 GiB HBM2}{+ 32 GiB DDR4} & \two{8 GiB HBM2}{+ 4.5 GiB DDR4} & \two{8 GiB}{HBM2} & \two{32 GiB HBM2e}{+ 32 GiB DDR4} \\
    Peak bandwidth (GB/s) & 936 & 1,555 / 1,935 & 460 + 38 & 460 + 21.3 & 316 & 819 + 25.6 \\
    \two{PCIe link}{(GB/s per dir.)} & \two{Gen4 x16}{31.5} & \two{Gen4 x16}{31.5} & \two{Gen3 x16}{15.75} & \two{Gen3 x16}{15.75} & \two{Gen3 x16}{15.75} & \two{Gen5 x8}{31.5} \\
    Process node & 8 nm & 7 nm & 16 nm & 16 nm & 16 nm & 7 nm \\
    LUT / FF & -- & -- & 1,304K / 2,607K & 1,304K / 2,607K & 872K / 1,743K & 2,574K / 5,148K \\
    \two{DSP / BRAM36}{/ URAM} & -- & -- & \two{9,024 / 2,016}{/ 960} & \two{9,024 / 2,016}{/ 960} & \two{5,952 / 1,344}{/ 640} & \two{10,848 / 3,741}{/ 1,925} \\
    \bottomrule
  \end{tabular}}
  \vspace{-11pt}
\end{table}

\newsavebox{\tiersbox}
\newsavebox{\ablationbox}
\begin{table}[t]
  \centering
  \caption{}
  \label{tab:tiers}\label{tab:ablation}
  \scriptsize
  \setlength{\tabcolsep}{1.5pt}
  \sbox{\tiersbox}{\renewcommand{\arraystretch}{1.93393}%
  \begin{tabular}[t]{@{}ccccccccc@{}}
    \toprule
    & \multicolumn{4}{c}{U280} & \multicolumn{2}{c}{RTX 3090} & \multicolumn{2}{c}{A100} \\
    \cmidrule(lr){2-5}\cmidrule(lr){6-7}\cmidrule(l){8-9}
    Design & HBM & DDR4 & Host & NVMe & GDDR & Host & HBM & Host \\
    \midrule
    Qwen3-Coder, FP8 & 5.3 & 13.7 & 11.8 & -- & 16.6 & 14.3 & 30.9 & -- \\
    Qwen3-Next, W4 & 3.4 & 29.9 & 10.6 & -- & 20.1 & 23.8 & 26.5 & 17.4 \\
    Qwen3-Next, FP8 & 2.4 & 33.5 & 23.9 & 20.6 & 19.4 & 61.0 & 31.0 & 49.4 \\
    \bottomrule
  \end{tabular}}%
  \sbox{\ablationbox}{\renewcommand{\arraystretch}{1.01855}\setlength{\aboverulesep}{1pt}\setlength{\belowrulesep}{1pt}%
  \begin{tabular}[t]{@{}cccc@{}}
    \toprule
    Design & Lanes (SLR0/1/2) & Est. & Tokens/s \\
    \midrule
    Earlier HLS-lane prototype & 22 (8/7/7) & -- & 44.0 \\
    + RTL lanes, Proxy Ports & 26 (17/9/0) & 66.3 & 66.9 \\
    + Estimates, no budgets & 31 (17/12/2) & 76.1 & not routed \\
    + Per-SLR budgets (full) & 30 (17/11/2) & 74.2 & \textbf{74.7} \\
    Full without Proxy Ports & 30 (17/11/2) & 65.7 & -- \\
    Full on 15 pseudo-channels & 30 (17/11/2) & 42.3 & -- \\
    \midrule
    Lane unit (LUTs / DSPs) & Optimized HLS & \multicolumn{2}{c}{RTL} \\
    \midrule
    GEMV lane & 7,777 / 94 & \multicolumn{2}{c}{3,875 / 87} \\
    GEMV and attention lane & 30,885 / 294 & \multicolumn{2}{c}{10,585 / 162} \\
    \bottomrule
  \end{tabular}}%
  \edef\tiersw{\the\dimexpr(\linewidth-6pt)*\numexpr\wd\tiersbox\relax/\numexpr\wd\tiersbox+\wd\ablationbox\relax\relax}%
  \edef\ablationw{\the\dimexpr\linewidth-6pt-\tiersw\relax}%
  \begin{minipage}[b]{\tiersw}\centering\footnotesize
  (a) Weights in each memory at the start of decoding (GB)
  \end{minipage}\hfill
  \begin{minipage}[b]{\ablationw}\centering\footnotesize
  (b) Co-design ablation and lane units, Llama-3.1-8B W4
  \end{minipage}\\[6pt]
  \resizebox{\tiersw}{!}{\raisebox{\depth}{\usebox{\tiersbox}}}\hfill
  \resizebox{\ablationw}{!}{\raisebox{\depth}{\usebox{\ablationbox}}}
  \vspace{-11pt}
\end{table}

\begin{table*}[t]
  \centering
  \caption{Decode throughput comparison.}
  \label{tab:results}
  \footnotesize
  \setlength{\tabcolsep}{2pt}
  \begin{minipage}[b]{0.63\textwidth}\centering\small (a) Throughput, card power, and energy efficiency of our designs and the GPUs at context length 512, decoding 256 tokens, and U280 throughput at eight context and decode lengths\end{minipage}\hfill
  \begin{minipage}[b]{0.355\textwidth}\centering\small (b) Published results of prior FPGA designs, each compared with our matching design\end{minipage}\\[3pt]
  \begin{minipage}[t]{0.63\textwidth}
  \centering
  \resizebox{\linewidth}{!}{%
  \begin{tabular}[t]{@{}ccccccccccccccccc@{}}
    \toprule
    & & \multicolumn{3}{c}{\textbf{U280 (ours)}} & \multicolumn{2}{c}{RTX 3090} & \multicolumn{2}{c}{A100} & \multicolumn{4}{c}{\textbf{U280 tok/s, context 512}} & \multicolumn{4}{c}{\textbf{U280 tok/s, context 1,024}} \\
    \cmidrule(lr){3-5}\cmidrule(lr){6-7}\cmidrule(lr){8-9}\cmidrule(lr){10-13}\cmidrule(lr){14-17}
    LLM & Format & tok/s & W & tok/J & tok/s & tok/J & tok/s & tok/J & 256 & 512 & 1,024 & 2,048 & 256 & 512 & 1,024 & 2,048 \\
    \midrule
    Llama-3.2-1B & W8 & 182.5 & 66.3 & \textbf{2.75} & 380.1 & 1.26 & \textbf{454.4} & 2.34 & 182.5 & 180.7 & 172.5 & 161.8 & 167.0 & 165.0 & 161.8 & 152.3 \\
    Llama-3.2-1B & W4 & 301.1 & 61.5 & \textbf{4.90} & 492.8 & 1.55 & \textbf{575.4} & 3.05 & 301.1 & 297.9 & 285.4 & 268.4 & 277.4 & 274.0 & 268.4 & 253.3 \\
    Llama-3.1-8B & W4 & 74.7 & 76.3 & \textbf{0.98} & 150.1 & 0.44 & \textbf{197.2} & 0.70 & 74.7 & 74.1 & 71.3 & 67.6 & 69.5 & 68.7 & 67.6 & 64.3 \\
    Granite-4.0-H-1B & W4 & 215.7 & 62.3 & \textbf{3.46} & 353.4 & 1.11 & \textbf{361.8} & 2.16 & 215.7 & 215.1 & 212.0 & 207.8 & 209.9 & 209.1 & 207.8 & 203.7 \\
    Qwen1.5-MoE-A2.7B & W4 & 174.9 & 68.6 & \textbf{2.55} & 270.2 & 0.81 & \textbf{300.6} & 1.51 & 174.9 & 173.1 & 166.6 & 157.3 & 162.4 & 160.5 & 157.3 & 143.3 \\
    Qwen3-Coder-30B-A3B & FP8 & 26.8 & 58.0 & 0.46 & 24.8 & 0.12 & \textbf{158.4} & \textbf{0.87} & 26.8 & 26.3 & 25.3 & 22.9 & 24.9 & 24.4 & 23.0 & 20.7 \\
    Qwen3-Next-80B-A3B & W4 & 29.3 & 65.2 & \textbf{0.45} & 39.4 & 0.17 & \textbf{49.0} & 0.36 & 29.3 & 29.1 & 28.5 & 28.0 & 29.1 & 29.2 & 28.6 & 27.9 \\
    Qwen3-Next-80B-A3B & FP8 & \textbf{24.5} & 64.3 & \textbf{0.38} & 16.0 & 0.11 & 22.0 & 0.19 & 24.5 & 24.4 & 22.9 & 21.4 & 22.2 & 20.6 & 22.1 & 22.2 \\
    BitNet b1.58-2B & ternary & 249.4 & 63.7 & \textbf{3.92} & 292.4 & 0.85 & \textbf{295.1} & 1.61 & 249.4 & 237.6 & 222.4 & 198.2 & 212.4 & 208.9 & 197.1 & 178.1 \\
    Ternary-Bonsai-27B & ternary & 32.7 & 62.6 & \textbf{0.52} & \textbf{65.2} & 0.19 & 59.7 & 0.21 & 32.7 & 32.5 & 32.0 & 30.9 & 31.6 & 31.4 & 30.9 & 30.0 \\
    \bottomrule
  \end{tabular}}%
  \end{minipage}\hfill
  \begin{minipage}[t]{0.355\textwidth}
  \centering
  \resizebox{\linewidth}{!}{%
  \begin{tabular}[t]{@{}ccccccccc@{}}
    \toprule
    & & \multicolumn{3}{c}{Prior design} & \multicolumn{4}{c}{\textbf{Ours}} \\
    \cmidrule(lr){3-5}\cmidrule(lr){6-9}
    Design, board & LLM, format & tok/s & W & tok/J & LLM, format & tok/s & W & tok/J \\
    \midrule
    \begin{tabular}[c]{@{}c@{}}LUT-LLM\\V80\end{tabular} & \begin{tabular}[c]{@{}c@{}}Qwen3-1.7B\\W2A3\end{tabular} & 196 & -- & -- & \begin{tabular}[c]{@{}c@{}}Llama-3.2-1B\\W4A8\end{tabular} & 301.1 & 61.5 & 4.90 \\
    \midrule
    \begin{tabular}[c]{@{}c@{}}FlexLLM\\U280\end{tabular} & \begin{tabular}[c]{@{}c@{}}Llama-3.2-1B\\W4A4KV8\end{tabular} & 144.7 & 67.0 & 2.16 & \begin{tabular}[c]{@{}c@{}}Llama-3.2-1B\\W4A8\end{tabular} & 301.1 & 61.5 & 4.90 \\
    \midrule
    \begin{tabular}[c]{@{}c@{}}FlightLLM\\U280\end{tabular} & \begin{tabular}[c]{@{}c@{}}LLaMA2-7B\\W3.5A8, sparse\end{tabular} & 55 & -- & -- & \begin{tabular}[c]{@{}c@{}}Llama-3.1-8B\\W4A8\end{tabular} & 74.7 & 76.3 & 0.98 \\
    \midrule
    \begin{tabular}[c]{@{}c@{}}EdgeLLM\\VCU128\end{tabular} & \begin{tabular}[c]{@{}c@{}}ChatGLM2-6B\\W4A16KV16\end{tabular} & 51.4 & 56.9 & 0.90 & \begin{tabular}[c]{@{}c@{}}Llama-3.1-8B\\W4A8\end{tabular} & 74.7 & 76.3 & 0.98 \\
    \specialrule{\heavyrulewidth}{\dimexpr\aboverulesep+\lightrulewidth-\heavyrulewidth\relax}{\belowrulesep}
    \begin{tabular}[c]{@{}c@{}}PENSA\\U50\end{tabular} & \begin{tabular}[c]{@{}c@{}}BitNet b1.58-2B\\ternary, A8\end{tabular} & 58.1 & -- & -- & \begin{tabular}[c]{@{}c@{}}BitNet b1.58-2B\\ternary, A8\end{tabular} & 249.4 & 63.7 & 3.92 \\
    \bottomrule
  \end{tabular}}%
  \end{minipage}
  \vspace{-11pt}
\end{table*}

\begin{table}[t]
  \centering
  \caption{Quantization and quality: Hugging Face checkpoint $\rightarrow$ ours.}
  \label{tab:scheme}
  \footnotesize
  \setlength{\tabcolsep}{3pt}
  \renewcommand{\arraystretch}{0.8999}
  \setlength{\aboverulesep}{1pt}\setlength{\belowrulesep}{1pt}
  \resizebox{\columnwidth}{!}{%
  \begin{tabular}{@{}lccccc@{}}
    \toprule
    LLM & W & A & KV (K / V) & WikiText-2 PPL & MMLU \\
    \midrule
    Llama-3.2-1B & W8 & A8 & bf16 / FP8 & 9.575 $\rightarrow$ 9.598 (+0.24\%) & 34.50 $\rightarrow$ 33.92 \\
    Llama-3.2-1B & W4 & A8 & bf16 / FP8 & 10.494 $\rightarrow$ 10.525 (+0.30\%) & 30.60 $\rightarrow$ 31.19 \\
    Llama-3.1-8B & W4 & A8 & bf16 / FP8 & 7.904 $\rightarrow$ 7.915 (+0.13\%) & 70.76 $\rightarrow$ 70.57 \\
    Granite-4.0-H-1B & W4 & A8 & bf16 / FP8 & 14.427 $\rightarrow$ 14.406 ($-$0.15\%) & 55.65 $\rightarrow$ 56.34 \\
    Qwen1.5-MoE-A2.7B & W4 & A8 & int8 / FP8 & 7.994 $\rightarrow$ 8.002 (+0.09\%) & 63.35 $\rightarrow$ 63.16 \\
    Qwen3-Coder-30B-A3B & FP8 & A8 & bf16 / FP8 & 9.227 $\rightarrow$ 9.214 ($-$0.13\%) & 78.85 $\rightarrow$ 78.56 \\
    Qwen3-Next-80B-A3B & W4 & A8 & bf16 / FP8 & 6.049 $\rightarrow$ 6.052 (+0.05\%) & 87.62 $\rightarrow$ 87.72 \\
    Qwen3-Next-80B-A3B & FP8 & A8 & bf16 / FP8 & 5.933 $\rightarrow$ 5.950 (+0.29\%) & 87.52 $\rightarrow$ 87.43 \\
    BitNet b1.58-2B & ternary & A8 & bf16 / FP8 & 14.187 $\rightarrow$ 14.216 (+0.20\%) & 55.26 $\rightarrow$ 55.56 \\
    Ternary-Bonsai-27B & ternary & A8 & bf16 / FP8 & 11.347 $\rightarrow$ 11.365 (+0.16\%) & 77.58 $\rightarrow$ 77.29 \\
    \bottomrule
  \end{tabular}}
  \vspace{-11pt}
\end{table}

\begin{figure}[t]
  \centering
  \includegraphics[width=0.91\linewidth]{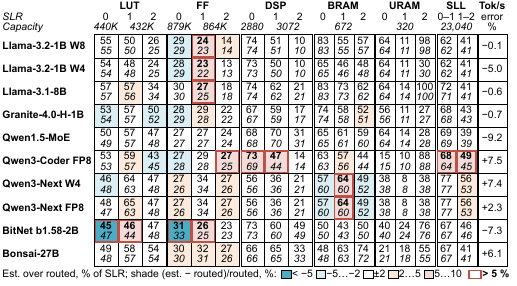}
  \caption{Per-SLR estimates against the ten routed designs and the board.}
  \label{fig:resources}
  \vspace{-11pt}
\end{figure}
\begin{figure}[t]
  \centering
  \includegraphics[width=\linewidth]{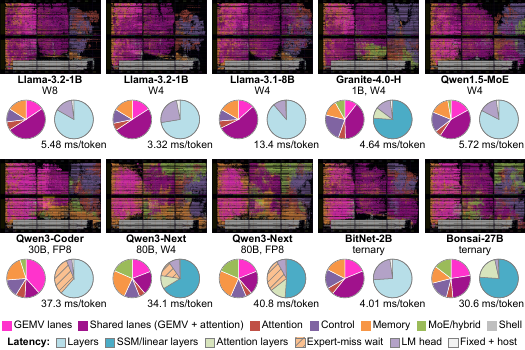}
  \caption{Placed and routed layouts on the U280, with LUT use by unit category (left pie) and per-token decode time (right pie). Shell: AMD's static platform region.}
  \label{fig:layouts}
  \vspace{-11pt}
\end{figure}

\textbf{Performance and energy efficiency.}\label{sec:eval-perf}
Across the eight context and decode lengths of Table~\ref{tab:results}(a),
each design's lowest throughput is 71\% to 95\% of its highest.
For prior hand-designed FPGA accelerators, we take their published results (Table~\ref{tab:results}(b)).
FlexLLM~\cite{zhang2026flexllm} runs the same model, Llama-3.2-1B,
on the same board at the same setting.
With 4-bit weights, our design decodes 301.1 tokens/s,
2.08$\times$ FlexLLM's 144.7,
and 2.04$\times$ to 2.09$\times$ at all eight of FlexLLM's context and decode lengths.
At the same HBM bandwidth,
our kernels run at 200~MHz against its 292~MHz.
FlightLLM~\cite{zeng2024flightllm} gives its rate without a stated context
and EdgeLLM~\cite{huang2025edgellm} at the 128th decoded token;
our 4-bit Llama-3.1-8B decodes 1.36$\times$ and 1.45$\times$ their rates, respectively.
LUT-LLM's released code~\cite{he2025lutllm} gives about 196 tokens/s,
from RTL co-simulation of one decoder block without the LM head.
Over these four designs,
ours reach 1.6$\times$ their throughput on average
(geometric mean).
For the extreme low-bit BitNet-2B,
PENSA~\cite{kalyanasundaram2026pensa} decodes up to 58.1 tokens/s on a U50;
ours decodes 249.4, about 4.3$\times$ its rate.
None of these five prior designs decodes MoE LLMs
or hybrid LLMs.

The U280 card draws 58 to 76~W,
and the two GPUs draw 117 to 348~W.
On average over the designs of Table~\ref{tab:results} (geometric mean),
our designs reach 3.0$\times$ and 1.5$\times$
the tokens per joule of the RTX 3090 and A100, respectively.
With 49\% of the RTX 3090's and 30\% of the A100's memory bandwidth
(Table~\ref{tab:platforms}),
our designs reach 70\% and 52\% of the GPUs' throughput (geometric mean).

\textbf{Tiered expert storage.}\label{sec:eval-tiers}
The four MoE designs differ in where their weights reside (Table~\ref{tab:tiers}(a)).
An expert resident in a faster memory also stays in its store in a slower one,
and the table counts it once, in the faster memory.
Qwen1.5-MoE resides entirely in HBM on the U280 and in GPU memory on both GPUs.
Qwen3-Coder resides entirely in the memory of the A100,
whereas its experts exceed the HBM of the U280 and the memory of the RTX 3090.
The RTX 3090 reads them from host DRAM over PCIe, which limits its rate;
the U280 reads them from on-board DDR4 and host memory,
and decodes faster than the RTX 3090.
The DDR4 bandwidth of the U280, 38~GB/s, is close to the 31.5~GB/s
of the GPUs' PCIe links (Table~\ref{tab:platforms}).
The experts of the 4-bit Qwen3-Next exceed the memory of all three platforms.
Those of the FP8 Qwen3-Next do as well,
and on the U280 its expert store extends through host memory to NVMe.
The Loaders read host memory directly over PCIe as bus masters,
and the U280 decodes the FP8 design faster than both GPUs.
Expert-miss waits take 32\% of the time of Qwen3-Coder
and 27\% of the FP8 Qwen3-Next,
against 13\% of the 4-bit Qwen3-Next (Figure~\ref{fig:layouts}).
On an earlier Qwen3-Coder design,
copying the next layer's experts early raises the share of selected experts resident in HBM
from 81\% to 92\% and the throughput by 19.9\% at contexts up to 512.

\textbf{Co-design ablation.}\label{sec:eval-guided}
Table~\ref{tab:ablation}(b) adds the parts of the co-design step by step
to the Llama-3.1-8B design with 4-bit weights,
starting from an earlier HLS-lane prototype that uses none of these parts.
It failed to route with 24 lanes
and was cut to 22 lanes, which decode 44.0 tokens/s.
The second row adds the RTL lanes and the Proxy Ports.
At the same parallelism and clock,
the RTL GEMV lane takes about half the LUTs of its optimized HLS form,
and the RTL shared lane about one third
(lower rows of Table~\ref{tab:ablation}(b)).
With the freed area, this design has 26 lanes, set without the model,
and decodes 66.9 tokens/s.
The last three designs of Table~\ref{tab:ablation}(b) share the RTL lanes,
controller, clock, and host program,
and differ only in the lanes of each SLR.
The third row sets the lanes by the estimated throughput alone,
without the per-SLR budgets of Eq.~\eqref{eq:budgets}:
the best candidate has 31 lanes and uses every HBM2 port.
In its link, SLR0 showed level-7 global and level-6 short congestion,
and routing started with 1.08 million node overlaps,
against 1,356 for the 30-lane design, and failed.
The last row adds the per-SLR budgets and is the full co-design.
Its 30 lanes route and decode 74.7 tokens/s,
11.6\% above the second row,
against an estimated gain of 12.0\%,
and 1.7$\times$ the rate of the HLS-lane prototype.
On the memory side, with HBM masters in SLR1 and SLR2 in place of Proxy Ports,
the HBM clock falls from 450 to 342~MHz,
and each HBM2 port delivers 10.9~GB/s instead of 12.6~GB/s;
the 30 lanes then drop from 74.2 to 65.7~tokens/s (Table~\ref{tab:ablation}(b)).
A lane in SLR1 costs 548~SLLs and 2.4K~LUTs for its proxy master and relays.
Two readers that share one pseudo-channel receive 6.6~GB/s each,
and the 30 lanes on 15 pseudo-channels reach 42.3~tokens/s (Table~\ref{tab:ablation}(b)).
At 12.6~GB/s per port, the 30 lanes of the full co-design read 378~GB/s,
82\% of the HBM2 peak, while they stream weights.
Each token runs 289 operators on these lanes.

\textbf{Configuration analysis and guided design.}\label{sec:eval-analysis}
The throughput estimates that guide each configuration differ
from the board by 4.6\% on average over the ten designs (Figure~\ref{fig:resources}),
largest for Qwen1.5-MoE, $-$9.2\%,
and by 6.4\% on average across all eight context and decode lengths of Table~\ref{tab:results}(a).
With every memory path at its nominal peak rate instead of its measured rate,
the mean error rises to 14.0\%,
and to 23\% to 35\% for the three designs with experts outside HBM.
For four resident expert sets of an earlier Qwen1.5-MoE design,
the estimates follow the board's order,
and their errors, $-$5.5\% to $-$4.2\%, differ by at most 1.3 points.
The per-SLR resource and SLL estimates are within 10\% of the routed designs,
with median errors of 1.3\% for LUTs, 2.7\% for FFs, and 0.5\% for SLLs.
The estimates also rank the changes after a failed check:
for a Qwen1.5-MoE configuration with a congested SLR0,
moving one CU to SLR2 left the estimated throughput unchanged,
while removing one SLR0 lane lowered it by 5.1\% to 6.6\%;
the design routed with the CU in SLR2.

\textbf{Layout, resource, and latency breakdown.}\label{sec:eval-layout}
The ten designs use 16 to 30 lanes, 10 to 17 of them in SLR0.
In each routed layout of Figure~\ref{fig:layouts}, HBM is on the left;
the HBM interfaces and most GEMV lanes sit in SLR0,
the shared lanes mostly in SLR1,
the controller and attention-only units in SLR2,
and the MoE and hybrid units mostly in SLR1 and SLR2.
Without the platform shell, the lanes take 39\% to 68\% of each design's LUTs,
and the MoE and hybrid units at most 19\%.
Without the expert-miss wait, the layers before the LM head take 61\% to 97\% of each token's time.

\section{Conclusion}\label{sec:conclusion}
High HBM bandwidth alone does not give high decode performance
on HBM-equipped FPGAs; compute, memory organization, and floorplan
jointly determine it.
Chimera co-designs the three in one configurable architecture.
Its designs
reach a geometric mean of 1.6$\times$ the throughput of
four prior state-of-the-art FPGA implementations
and 3.0$\times$ and 1.5$\times$ the energy efficiency
of the RTX 3090 and A100.

\bibliographystyle{IEEEtran}
\bibliography{IEEEtran_ctl,refs}

\end{document}